\documentclass[11pt,notitlepage,a4paper]{article}
\pdfoutput=1
\usepackage[a4paper,left=3cm,right=3cm,top=3cm,bottom=3cm,footskip=1.5cm]{geometry}
\usepackage[bf]{titlesec}
\usepackage{tocloft}

\newlength{\mylen}	
\usepackage[T1]{fontenc}
\usepackage[english]{babel}
\usepackage{amsmath,amssymb,amscd,amsfonts}
\usepackage{mathtools}
\usepackage{url}
\usepackage{tikz} 
\usepackage{enumitem}
\usepackage[hidelinks]{hyperref}
\hypersetup{
    pdftitle={A pragmatic approach to formal fundamental physics},
    pdfauthor={Daniel Friedan},
}
\usepackage[open,
  openlevel=2,
  atend,
  numbered]{bookmark}
\def\eq{\begin{equation}}
\def\en{\end{equation}}
\def\eqa{\begin{eqnarray}}
\def\ena{\end{eqnarray}}
\def\ket#1{|#1\rangle}
\def\bra#1{\langle #1 |}
\def\Reals{\mathbb{R}}
\def\Integers{\mathbb{Z}}
\def\expval#1{\langle \, #1 \,\rangle}
\begin{document}
%
%
%
\thispagestyle{empty}
\begin{center}
{\LARGE A pragmatic approach to formal fundamental physics}\\[7ex]
{\large Daniel Friedan}\\[2ex]
New High Energy Theory Center, Rutgers University\\
and Natural Science Institute, The University of Iceland\\
dfriedan@gmail.com\\[2ex]
October 22, 2018
\end{center}

\vspace*{4ex}

{\centering
\large\bfseries Abstract
\vskip2ex
}

A minimal practical formal structure for a fundamental theory is
suggested.  A mechanism that produces such a structure is reviewed.
The proposed mechanism has possibilities of producing non-canonical
phenomena in SU(2) and SU(3) quantum gauge theories.  These might
provide testable conditional predictions.  One possibility is a
vacuum condensate of SU(2) gauge fields
derived from certain trajectories of the SU(2) Yang-Mills flow.

\setcounter{tocdepth}{2}
\tableofcontents
\newpage
%
%
\section{Formal fundamental physics}

For 45 years, the most fundamental theory of physics has been the
Standard Model combined with General Relativity.
It describes almost everything known at distances larger
than about $(10^{3} \text{\small GeV})^{-1}$.
Dark matter, neutrino mixing, and some CP violation are the only
observed phenomena left unexplained.
The project of formal fundamental physics is to hypothesize a more
comprehensive formal machinery that includes the SM+GR and 
makes predictions that can be checked against experiment.
It seems overreaching
to attempt to predict
the SM+GR itself.
Sufficient would be
conditional predictions of the form
\emph{if the formalism
produces the SM+GR then it must also produce such and such testable phenomena
beyond the SM+GR}.
An example is proton decay conditionally predicted by
Grand Unification.
If a conditional prediction were to check successfully
against experiment,
then the proposed formalism would become a serious candidate for a more 
fundamental theory.

SM+GR is an {effective} quantum field theory with
short distance cutoff on the order of $(10^{3} \text{\small
GeV})^{-1}= 10^{16} \ell_{P}$
where $\ell_{P}$ is the Planck distance.
GR can be treated as an effective QFT
because
quantum effects are negligible at distances
${\gg}\,\ell_{P}$.
The effective QFT
with short distance cutoff $10^{16}\ell_{P}$ 
is indistinguishable from classical GR.
A more comprehensive formalism should be capable of producing an 
effective QFT such as the SM+GR and 
should make definite conditional predictions
of observable phenomena beyond the SM+GR.

In the 45 years since the SM was verified, none of the attempts at
formal fundamental physics
have worked (in the strict theoretical physics sense of `worked' --- making predictions
that check successfully against experiment).
One reaction is to give up 
on the project, perhaps hoping that experiment will
eventually provide more guidance.  
Alternatively,
it might be useful
to reexamine the assumptions that have
guided the formal fundamental physics enterprise
and reconsider paths not taken.
SM+GR already encodes very much
experimental evidence.
New high energy discoveries will most likely
lead to a new effective QFT that improves incrementally on the SM.
The question for formal fundamental physics
is how to use a specific effective QFT such as the SM+GR
as guidance towards a formalism more comprehensive and more predictive than effective 
QFT in general.

\subsection{Against Quantum Gravity}

It might be useful to question the truism
that General Relativity and Quantum Mechanics have to be reconciled
in a theory of Quantum Gravity.
On the contrary, 
there is no appreciable conflict
at distances ${\gg}\,{\ell_{P}}$.
The smallest distance presently accessible to experiment
is roughly
$L_{\text{exp}} = (10^{3} \text{\small GeV})^{-1} =10^{16} 
\ell_{P}$.
To suppose a conflict is to extrapolate
the validity of both Quantum Mechanics and General Relativity
over 16 orders of magnitude
from $10^{16} \ell_{P}$
down to $\ell_{P}$.
Nothing is known about physics 
at such small distances.
\nobreak
\begin{center}
\begin{tikzpicture}[x=0.275cm]
\draw (0,0) -- (32,0);
\foreach \j in {1,4,...,31} 
{
\draw (\j,0.1) --  + (0,-0.2) + (0,+0.2) ;
}
\foreach[evaluate={\i=int(19-\j)}] \j in {1,4,...,31} 
{
\draw (\j,+0.3) node[above] {\scriptsize$10^{\i}$};
\draw (\j,-0.3) node[below] {\scriptsize$10^{\j}$};
}
\draw (-4.0,+0.5) node[above]{\small energy} ;
\draw (-4.0,+0.1) node[above]{\small (GeV)} ;
\draw (-4.0,-0.1) node[below]{\small distance} ;
\draw (-4.0,-0.5) node[below]{\small (${\ell_{P}}$)} ;
\draw (34.5,+0.0) node {$\cdots\cdots$};
\draw(0,-1.3) node[below]{\small ${\ell_{P}}$};
\draw(16,-1.3) node[below]{\small $L_{\text{exp}}$};
\draw(31.8,-1.3) node[below]{\small $1{\text{mm}}$};
\draw(23,-2.3) node[below]{\small Quantum Mechanics};
\draw[->](16,-2.27) --  (36,-2.27);
\draw[dashed] (0,-2.27) --  (15,-2.27);
\draw(13,-2.32) node[below]{\footnotesize ?};
\draw(9.5,-2.32) node[below]{\footnotesize ??};
\draw(5.5,-2.32) node[below]{\footnotesize ???};
\draw(1,-2.32) node[below]{\footnotesize ????};
\end{tikzpicture}
\end{center}
Such a presumptuous extrapolation beyond
the physical evidence
would be justifiable if it yielded a testable prediction,
as for example the extrapolation of Grand Unification gave
the conditional prediction of proton decay.
But it is implausible that any proposed theory of Quantum Gravity
can be checked experimentally
given that the smallest distance presently accessible to experiment
is $10^{16} \ell_{P}$.
There is no practical possibility of checking whether any proposed 
theory of Quantum Gravity actually describes the real world.
Without the possibility of an experimental test,
except in fantasy,
any such extreme extrapolation beyond the experimental evidence 
is unlikely to be useful for fundamental physics.

\subsection{Against mathematical idealizations}

Formal structures are used in physics for practical purposes,
not as ideal mathematical forms.
A quantum field theory is used as an {\it effective} theory
describing physics at distances greater than some UV
cutoff at the short distance limit of the evidence.
An effective QFT says nothing about distances smaller than 
the UV cutoff.
It does not even suppose the existence of a space-time continuum.
Continuum QFT is a mathematical idealization
which extrapolates far beyond the practical use of the formalism.

Likewise, an S-matrix is used as an {\it effective} theory
that describes physics at distances smaller than 
the scattering region.
The asymptotic S-matrix
is a mathematical idealization
which
supposes ingoing scattering states produced infinitely early in time
and infinitely far from the scattering region
and outgoing scattering states detected infinitely later in time 
and infinitely far away.
Actual scattering experiments take place within a finite region of space 
over a finite period of time.
Again,
the idealized asymptotic S-matrix extrapolates far beyond
the practical use of the formalism.

These mathematical idealizations serve mathematical purposes.
But the practical limits of physical knowledge 
are encoded in 
the {\it effective} QFT with a UV cutoff and the {\it effective} S-matrix
with an IR cutoff.

\section{A minimal practical formal structure}

\subsection{An effective QFT for 
distances ${>}\,L$ and
an effective S-matrix for distances ${<}\,L$,
for observers at every distance scale $L\gg\ell_{P}$}

A leading edge high energy experiment
of size $L$
probes for new physics at distances ${<}\,L$.
The observer has in hand
an effective QFT
for distances $\gtrsim\! L$.
Short distance physics is probed
by sending things in
and measuring what comes out.
Measurements
are expressed as scattering amplitudes 
between states
of the effective QFT.
There is only
an effective S-matrix
with IR cutoff $L$
for short distance physics.

Prudence and practicality suggest a formalism
that corresponds with what is observable.
For every $L\,{\gg}\,\ell_{P}$ there should be
an effective QFT($L$) with UV cutoff $L$ and
an effective S-matrix($L$) with IR cutoff $ L$.
The meaning of ``short distance physics'' depends on the scale $L$
of the observer.
$L$ is a sliding distance scale.
The condition $L\,{\gg}\,\ell_{P}$
expresses the impracticality of experimenting
anywhere near the Planck scale.
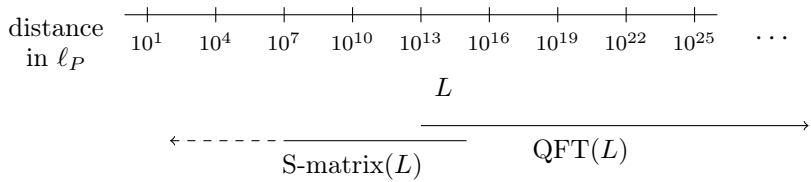
\begin{center}
\begin{tikzpicture}[x=0.3cm]
\draw (0,0) -- (26,0);
\foreach \j in {1,4,...,25} 
{
\draw (\j,0.1) --  + (0,-0.2) + (0,+0.2) ;
}
\foreach[evaluate={\i=int(19-\j)}] \j in {1,4,...,25} 
{
\draw (\j,-0.1) node[below] {\scriptsize $10^{\j}$};
}
\draw (-3,0.1) node[below]{\small distance} ;
\draw (-3,-0.3) node[below]{\small in ${\ell_{P}}$} ;
\draw (28.5,-0.1) node[below] {$\cdots$};
\draw(14,-0.7) node[below]{\small $L$};
\draw[->](13,-1.47) --  (30,-1.47);
\draw(20,-1.5) node[below]{\small QFT($L$)};
\draw(7,-1.67) -- (15,-1.67) ; 
\draw[<-,dashed] (2,-1.67) -- (7,-1.67) ; 
\draw(10,-1.7) node[below]{\small S-matrix($L$)};
\end{tikzpicture}
\end{center}
\subsection{QFT renormalization group operates from smaller distance $L$ 
to larger; S-matrix renormalization group operates from larger $L$ to smaller}
\label{sect:consistency}

The descriptions of physics must be consistent
as progress pushes to shorter distance $L'$.
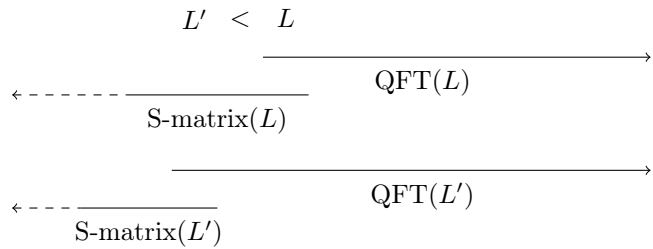
\begin{center}
\begin{tikzpicture}[x=0.3cm]
\draw(0,-1.3) node[below]{};
\draw(14,-1.3) node[below]{\small $L$};
\draw(12,-1.33) node[below]{\small $<$};
\draw(10,-1.3) node[below]{\small $L'$};
\draw[->](13,-2.07) --  (30,-2.07);
\draw(20,-2.1) node[below]{\small QFT($L$)};
\draw(7,-2.57) -- (15,-2.57) ; 
\draw[<-,dashed] (2,-2.57) -- (7,-2.57) ; 
\draw(11,-2.6) node[below]{\small S-matrix($L$)};
\draw[->](9,-3.57) --  (30,-3.57);
\draw(20,-3.6) node[below]{\small QFT($L'$)};
\draw(5,-4.07) -- (11,-4.07) ; 
\draw[<-,dashed] (2,-4.07) -- (5,-4.07) ; 
\draw(8,-4.1) node[below]{\small S-matrix($L'$)};
\end{tikzpicture}
\end{center}
\vspace*{-3ex}
\noindent
\begin{itemize}[labelindent=1.5em,labelsep=1em,leftmargin=*]
\item[{\bf C1}]
QFT($L$) must derive from QFT($L'$)
by the QFT renormalization group.
\item[{\bf C2}]
S-matrix($L$) must agree with
the scattering amplitudes derived from QFT($L'$)
at intermediate distances between $L'$ and $L$.
\item[{\bf C3}]
S-matrix($L'$) must derive
from S-matrix($L$) 
by the ``S-matrix renormalization group''.
\end{itemize}
\noindent
The S-matrix RG is the operation on effective S-matrices
that takes an effective S-matrix with IR cutoff distance $L$
to an effective S-matrix with smaller IR cutoff distance $L'$
by
using the scattering states at scale $L$
to make the scattering states
at the smaller scale $L'$.
The S-matrix RG and the QFT RG operate
in opposite directions on the distance scale.
The ideas of the effective S-matrix and the S-matrix RG 
are illustrated by the constructions
in section~\ref{sect:effectivestringSmatrix} and section~\ref{sect:smallhandlelogdivergence}
below.
Technical definitions remain to be formulated.

\subsection{An S-matrix does not imply a hamiltonian}

The formal structure QFT($L$)\,+\,S-matrix($L$) is local in $L$.  
An observer at scale $L$ makes only a modest extrapolation
by supposing
there will be a somewhat more fundamental effective QFT($L'$)
at somewhat smaller distance $L'$.
There is no presumption of QFT or any
quantum mechanical hamiltonian all the way down to $\ell_{P}$.
An S-matrix does not necessarily
come from a
microscopic hamiltonian.
An S-matrix can be derived
from  a microscopic hamiltonian,
but not vice versa.

Nor does  an S-matrix imply an effective QFT in the IR, even if
the IR limit of the S-matrix matches the scattering amplitudes of the effective QFT.
Such a coincidence only means that the effective QFT is consistent 
with the S-matrix.
There is still need for a mechanism that actually \emph{produces} the effective QFT.
One possible mechanism is the RG acting on a microscopic 
QFT or other microscopic hamiltonian system.
But there could be mechanisms for producing effective QFT that 
do not depend on such an extrapolation.


\section{A mechanism that produces such a formal structure}

A search for a mechanism that would produce a realistic QFT 
began with \cite{Friedan1980a,Friedan1980ThLBLAnnals}.
A mechanism was finally proposed in \cite{Friedan2003a}.
The mechanism produces a formal structure
such as described above.
The line of thought is sketched in
the Appendix.

\subsection{Summary}
\begin{enumerate}[labelindent=0em, leftmargin=*,widest=10]
\item
String theory provides a  way to construct 
a self-consistent S-matrix for short distance physics
without using a microscopic QFT.
\item
When the string worldsheet is an effective 2d-QFT with 2d 
UV cutoff distance $\Lambda^{-1}$,
the string S-matrix 
is an effective S-matrix($L$) with IR cutoff distance $L$
given in dimensionless units by
$
L^{2} = \ln\left(\Lambda/\mu\right)
$
where $\mu^{2}|dz|^{2}$ is the worldsheet metric.
The condition $L\,{\gg}\,1$ is the requirement $\Lambda^{-1}\,{\ll}\,\mu^{-1}$, 
i.e., the requirement that the 2d cutoff distance $\Lambda^{-1}$ be insignificant 
at the 2d distance scale $\mu^{-1}$
of the S-matrix calculation.
\item
The string background is encoded in the local worldsheet physics
at 2d distance $\Lambda^{-1}$.
\item
The S-matrix RG acts on S-matrix($L$) by integrating out the froth of small 
handles in the worldsheet,
increasing the 2d UV cutoff $\Lambda^{-1}$,
decreasing the IR cutoff $L$.
\item 
The effects of the froth of small handles are
replicated by a certain 
2d nonlinear model (2d-NLM) 
called the \emph{$\lambda$-model}.
The $\lambda$-model is mathematically natural.
The target manifold is the space of 
effective 2d-QFTs of the worldsheet,
parametrized by the
effective 2d coupling constants $\lambda^{i}(\Lambda)$
at 2d scale $\Lambda^{-1}$.
These $\lambda^{i}(\Lambda)$
are the modes of the classical background space-time
fields with UV cutoff $L$.
The target manifold  of the $\lambda$-model
is thus the space of
classical space-time fields with UV cutoff $L$.
\item
The froth of small handles
is replaced by the $\lambda$-fluctuations
at 2d distances ${<}\,\Lambda^{-1}$,
which produce
an effective worldsheet QFT with 2d UV cutoff $\Lambda^{-1}$.
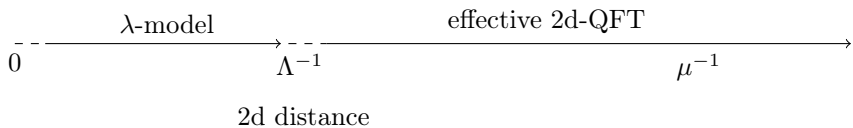
\begin{center}
\begin{tikzpicture}
\draw[dashed](0,0) -- (0.35,0);
\draw[->](0.4,0) -- (3.5,0);
\draw[dashed](3.6,0) -- (4.1,0);
\draw[->](4.1,0) -- (11,0);
\draw(0,0) node[below]{\small 0};
\draw(2,0) node[above]{\small$\lambda$-model};
\draw(3.75,0) node[below]{\small$\Lambda^{-1}$};
\draw(7,0) node[above]{\small effective 2d-QFT};
\draw(9,0) node[below]{\small$\mu^{-1}$};
\draw(3.8,-0.7) node[below]{\small 2d distance};
\end{tikzpicture}
\end{center}
\item
Integrating out the $\lambda$-fluctuations has the same effect as
integrating out the froth of small handles
so the 2d-RG of the $\lambda$-model
implements the S-matrix RG.
\item
The 2d-RG of the $\lambda$-model also produces
a measure on the target manifold.
This is the \emph{a priori} measure
of the 2d-NLM.
A measure on the target manifold of the $\lambda$-model
is a functional integral over the
space-time fields with UV cutoff $L$,
i.e., an
effective quantum
field theory QFT($L$).
\item
The quantum states of QFT($L$)
are the quantum string backgrounds.
\item
QFT($L$) and S-matrix($L$) automatically satisfy the consistency 
conditions {\bf C1}, {\bf C2}, {\bf C3}.
The $\lambda$-model produces a consistent realization of the minimal practical
formal structure described above.
\item
The effective 
QFT($L$) is produced by a 2d mechanism that does not necessarily correspond
to canonical quantization (except perturbatively).
There are concrete possibilities of
nonperturbative semi-classical 2d effects 
which could
produce non-canonical degrees of freedom and 
non-canonical interactions in QFT($L$).
These 2d effects are the 2d winding modes and 2d instantons coming from
nontrivial homotopy groups $\pi_{1}$ and $\pi_{2}$ of the space of space-time fields
which is the target 
manifold of the $\lambda$-model.
These homotopy groups are nontrivial
when the space-time fields include SU(2) 
or SU(3) gauge fields in four space-time dimensions.
\end{enumerate}
The last point is
the main reason for investigating the $\lambda$-model 
as a formalism for fundamental physics.
There are possibilities
of conditional predictions
of the form
{\it if QFT($L$) contains SM+GR, then
it predicts certain specific
non-canonical degrees of freedom and 
interactions
beyond those of the canonically quantized
effective quantum field theory.}

\subsection{2d-QFT of the string worldsheet}
\label{sect:2dQFT}

In the general renormalizable 2d nonlinear model
\eq
\int  e^{-\int d^{2}z \; g_{\mu\nu}(X) 
\partial X^{\mu} \bar \partial X^{\nu}}
\mathcal{D} X
\qquad
X(z) \in M
\en
the field $X(z)$ takes values in a target manifold $M$.
The 2d coupling constants  are given 
by a Riemannian metric $g_{\mu\nu}(X)$ on $M$.
The manifold $M$ is taken compact and $g_{\mu\nu}(X)$
euclidean signature
so that the 2d-QFT will be well defined.
The 2d-RG
\eq
\Lambda \frac{\partial}{\partial \Lambda}\, g_{\mu\nu}(X)  = - 
R_{\mu\nu}(X) +  O(R^{2})
\en
drives the 2d-NLM to a solution of $R_{\mu\nu}=0$.

The 2d-QFT of the string worldsheet is an
elaboration of the general
2d-NLM in which the target manifold $M$ is space-time and the 2d coupling
constants consist of the space-time metric $g_{\mu\nu}(X)$ 
and also some non-abelian
gauge fields, scalar fields, fermion fields, etc.\ on the space-time
$M$.  The equation $\beta=0$ generalizing
$R_{\mu\nu}=0$ is a semi-realistic supersymmetric classical field equation which
includes GR and potentially the SM.

In abstract language,\\[1ex]
\hspace*{2em}
\begin{tabular}{rcl}
$\lambda^{i}$ &=& the 2d coupling constants,
\\[0.5ex]
$\phi_{i}(z)$ &=& the corresponding spin-0 scaling fields of the 
2d-QFT,
\\[0.5ex]
$\ket{\phi_{i}}$ &=& the corresponding radial quantization
states on the unit circle in 2d,
\\[0.5ex]
$G_{ij}$ &=& the natural 
metric $\langle\phi_{i}|\phi_{j}\rangle$.
\end{tabular}\\[1ex]
The index $i$ labels the modes of the space-time fields.
For example, the modes
of the space-time metric 
and the corresponding 2d fields are
\eq
\delta_{i} g_{\mu\nu}(X)=e^{ip_{\mu}(i)X^{\mu}}
h_{\mu\nu}{(i)}
\qquad
\phi_{i}(z) = 
e^{ip_{\mu}(i)X^{\mu}(z)}
h_{\mu\nu}{(i)}\: \partial 
X^{\mu}\bar \partial X^{\nu}
\en
The $\lambda^{i}$
form a system of local coordinates on the space of 
2d-QFTs.
The nearby 2d-QFTs are given by 
inserting in the worldsheet
\eq
e^{\int d^{2}z\, \lambda^{i}\phi_{i}(z) }
\en
The 2d scaling-dimensions are
\eq
\dim(\phi_{i})  = 2 +\delta({i})
\qquad
\dim(\lambda^{i}) = - \delta({i})
\qquad
\delta({i}) = p(i)^{2}
\en
The 2d-RG 
\eq
\Lambda \frac{\partial}{\partial \Lambda}\,\lambda^{i} = 
\beta^{i}(\lambda)
\qquad
\beta^{i}(\lambda) = -\delta(i) \lambda^{i} + O(\lambda^{2})
\en
drives the worldsheet 2d-QFT towards the $\beta=0$ submanifold
which is 
parametrized by the marginal coupling 
constants
\eq
\dim(\lambda^{i}) = - \delta({i})= - p(i)^{2}= 0
\en
which correspond to the zero-modes of the space-time fields.
There are no relevant operators,
no $\lambda^{i}$ with $\delta(i)<0$.
(There are no tachyons in the string S-matrix.)
The $\beta=0$ submanifold is stable under
the 2d-RG.
There are no unstable directions.
\subsection{Effective string S-matrix with IR cutoff $L$}
\label{sect:effectivestringSmatrix}

Let $\mu^{2}|dz|^{2}$ be the worldsheet metric.
Impose a 2d UV cutoff $\Lambda^{-1} \ll \mu^{-1}$.
The cutoff string propagator (the cutoff integral over 2d-cylinders) is
\eq
\int\limits_{0}^{\ln \left({\Lambda}/{\mu}\right)}
\bigg(
\sum_{i,j}  \;\; \ket{\phi_{i}}
\;G^{ij}e^{-\tau \delta(i)}
\bra{\phi_{j}}
\bigg)\,d\tau
=
\sum_{i,j}  \;\; \ket{\phi_{i}}\;\;
\frac{1-e^{-L^{2}\delta(i)}}{\delta(i)}
\,G^{ij}
\;\;\bra{\phi_{j}}
\en
where
\eq
L^{2} = \ln \left({\Lambda}/{\mu}\right)
\qquad
e^{-L^{2}\delta(i)}
=
\left({\Lambda}/{\mu}\right)^{-\delta(i)}
\en
The only modes that propagate are those that satisfy
\eq
\delta(i) > L^{-2}
\qquad\text{which is}
\qquad
p(i)^{2} > L^{-2}
\en
so the 2d UV cutoff $\Lambda^{-1}$ puts an IR cutoff $L$
on the string S-matrix.
An effective 2d-QFT of the worldsheet gives an 
effective string S-matrix($L$)
with $L$ given by
$L^{2} = \ln \left(\Lambda/\mu\right)$.

\subsection{Effective 2d coupling constants $\lambda^{i}(\Lambda)$}
\label{sect:effective2dcouplingconstants}
The effects of the 2d coupling constants
$\lambda^{i}({\Lambda})$
at 2d scale $\Lambda^{-1}$
are suppressed by the 
2d-RG running from $\Lambda^{-1}$ up to $\mu^{-1}$
\eq
\lambda^{i}(\mu) = \left({\Lambda}/{\mu}\right)^{-\delta(i)} 
\lambda^{i}({\Lambda})
= e^{-L^{2}\delta(i)}
\lambda^{i}({\Lambda})
\en
If $L^{2}\delta(i) > 1$
then $\lambda^{i}({\Lambda})$ is effectively irrelevant;
its effects on the worldsheet are negligible.
The only $\lambda^{i}(\Lambda)$ that matter
are the effectively marginal couplings
\eq
\delta(i) < L^{-2}
\qquad
\text{which is}\qquad p(i)^{2}< L^{-2}
\en
so there is a  UV cutoff distance $L$
on the modes of the space-time fields 
that are the coupling constants of the effective 2d-QFT of the worldsheet.

The 2d UV cutoff $\Lambda^{-1}$ 
separates the 2d coupling constants $\lambda^{i}$ into two subsets.
The $\lambda^{i}$ with $\delta(i)>L^{-2}$ are effectively irrelevant.
The corresponding $\phi_{i}(z)$ are the vertex operators 
for the propagating modes in the effective string S-matrix.
The $\lambda^{i}$ with $\delta(i)<L^{-2}$ are the effectively 
marginal coupling constants.
These are not exact solutions of $\beta=0$.
They are the off-shell
classical string backgrounds at distances ${>}\,L$.
The off-shell classical backgrounds
are prerequisites for quantum backgrounds at scales ${>}\,L$.

\subsection{Implement the S-matrix renormalization group}
\label{sect:smallhandlelogdivergence}

Consider the effect of a small handle in the worldsheet.
A small handle is made by identifying the boundaries of two holes 
of radius $r$ around two
points $z_{1}$ and $z_{2}$ which are close together in the worldsheet.
The identification is
\eq
z\leftrightarrow z'
\qquad
(z-z_{1})(z'-z_{2}) =  r^{2} e^{i\theta}
\en
Insert a sum over radial quantization states on each boundary circle.
Integrate over 
the moduli $z_{1},\,z_{2}, \, r,\,\theta$.
The integral over $\theta$ projects on the spin-0 states.
The effect of the small handle becomes the bi-local insertion
\eq
\frac12
\sum_{i_{1},i_{2}}
\int d^{2}z_{1}\, \phi_{i_{1}}(z_{1}) 
\int d^{2}z_{2}\, \phi_{i_{2}}(z_{2}) 
\int\limits_{\Lambda^{-1}}^{\frac12|z_{1}-z_{2}|}
\!\!\!\!\!dr
\;
r^{-1-\delta(i_{1})-\delta(i_{2})} 
\;
g_{\mathrm{str}}^{2}
G^{i_{1}i_{2}}
\en
where $G_{i_{1}i_{2}}$ is the natural metric on the space of 2d-QFTs
and $g_{\mathrm{str}}$ is the string coupling constant.
The integration region of interest here is 
$\Lambda^{-1}\,{<}\,|z_{1}-z_{2}|\,{\ll}\,\mu^{-1}$.
This is
where the small handle contributes to the 
local worldsheet physics.

The cutoff dependent contribution comes from the effectively marginal fields
\eq
\label{eq:handlecutoffdependence}
\frac12
\sum_{\,\delta(i_{1})\sim 0}
\;
\sum_{\delta(i_{2})\sim 0}
\;
\int d^{2}z_{1}\,\phi_{i_{1}}(z_{1}) 
\int d^{2}z_{2}\,\phi_{i_{2}}(z_{2}) 
\;g_{\mathrm{str}}^{2} G^{i_{1}i_{2}} \ln \left ( \Lambda |z_{1}-z_{2}|\right )
\en
The cutoff dependence of the small handle  
expressed by (\ref{eq:handlecutoffdependence}) can be canceled
by letting the effectively marginal 2d coupling constants $\lambda^{i}$
fluctuate locally on the worldsheet.
Make the $\lambda^{i}$  into sources $\lambda^{i}(z)$
so the worldsheet insertion becomes
\eq
e^{\int d^{2}z\, \lambda^{i}(z)\phi_{i}(z) } 
\en
Then set the $\lambda^{i}(z)$  fluctuating with 2-point correlation function
\eq
\label{eq:gaussianfluctuations}
\expval{\lambda^{i_{1}}(z_{1})
\;
\lambda^{i_{2}}(z_{2})
}
=
-g_{\mathrm{str}}^{2} G^{i_{1}i_{2}}  \ln \left ( \Lambda |z_{1}-z_{2}|\right )
\en
The cancellation of the single small handle 
(\ref{eq:handlecutoffdependence}) by the gaussian 
$\lambda$-fluctuations (\ref{eq:gaussianfluctuations})
holds in every background 2d-QFT.
Therefore the cutoff dependence of the entire nongaussian froth of small 
handles is canceled by $\lambda$-fluctuations
governed by the 2d-NLM
\eq
\label{eq:lambdamodel}
\int
e^{-\int d^{2}z\; {g_{\mathrm{str}}^{-2}}G_{ij}(\lambda) \partial \lambda^{i}\bar \partial 
\lambda^{j}}
\:\:
e^{\int d^{2}z\,\lambda^{i}(z)\phi_{i}(z)}
\;\mathcal{D}\lambda 
\qquad
\lambda(z) \in \mathcal{M}
\en
This 2d-NLM is the $\lambda$-model.
The target manifold is
\begin{center}
\begin{tabular}{r@{\hskip0.5em}c@{\hskip0.6em}l}
$\mathcal{M}$ 
& = &
the manifold of effective worldsheet 2d-QFTs
with 2d cutoff $\Lambda^{-1}$
\\[1ex]
& = &
the space of classical space-time fields with UV cutoff 
$L$ given by $L^{2} = \ln \left( {\Lambda}/{\mu}\right)$
\end{tabular}
\end{center}

The sum over handles
does not depend on the arbitrary choice of $\Lambda^{-1}$
so the sum over $\lambda$-fluctuations
at 2d distances ${<}\,\Lambda^{-1}$
has the same effect as the sum over
small handles,
since either of these sums
cancels the $\Lambda^{-1}$ dependence
of the sum over handles at 2d distances ${>}\,\Lambda^{-1}$.
Therefore integrating out the $\lambda$-fluctuations
at 2d distances ${<}\,\Lambda^{-1}$
is equivalent to integrating out the froth of small handles.
Integrating out the $\lambda$-fluctuations at 2d distances $< \Lambda^{-1}$
produces an effective 2d-QFT with UV cutoff $\Lambda^{-1}$
which gives
an effective string S-matrix($L$) with IR cutoff $L$.
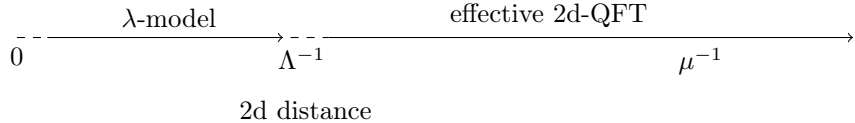
\begin{center}
\begin{tikzpicture}
\draw[dashed](0,0) -- (0.35,0);
\draw[->](0.4,0) -- (3.5,0);
\draw[dashed](3.6,0) -- (4.1,0);
\draw[->](4.1,0) -- (11,0);
\draw(0,0) node[below]{\small 0};
\draw(2,0) node[above]{\small$\lambda$-model};
\draw(3.75,0) node[below]{\small$\Lambda^{-1}$};
\draw(7,0) node[above]{\small effective 2d-QFT};
\draw(9,0) node[below]{\small$\mu^{-1}$};
\draw(3.8,-0.7) node[below]{\small 2d distance};
\end{tikzpicture}
\end{center}
The 2d-RG of the $\lambda$-model  operates
from smaller 2d distance $\Lambda'^{-1}$ up to
larger 2d distance $\Lambda^{-1}$
by integrating out the $\lambda$-fluctuations at 2d distances
between $\Lambda'^{-1}$ and $\Lambda^{-1}$.
This is equivalent to 
integrating out the small handles,
taking the effective S-matrix($L'$) with larger IR cutoff $L'$
to the effective S-matrix($L$) with smaller IR cutoff $L$.
Thus the 2d-RG of the $\lambda$-model
implements the S-matrix RG.

The ``froth of small handles'' has only a 
perturbative meaning
in string theory.
The $\lambda$-model replicates the perturbative froth
and is a nonperturbatively well defined 2d-NLM.
So the $\lambda$-model gives nonperturbative meaning to the froth of 
small handles.

\subsection{Production of an effective QFT with UV cutoff $L$}
\label{sect:productionofeffectiveQFT}

Like any 2d-NLM, the $\lambda$-model  is specified by two pieces of data
\vskip1ex

\begin{itemize}
\item the metric
$g_{\mathrm{str}}^{-2}G_{ij}(\lambda)$ on the target manifold $\mathcal{M}$
\vskip2ex

\item
the \emph{a priori} measure $\rho(\lambda) d\lambda$ on the target 
manifold $\mathcal{M}$
from which comes
the functional volume element 
in the functional integral (\ref{eq:lambdamodel})
\eq
\mathcal{D}\lambda = \prod_{z} 
\rho(\lambda(z))\,d\lambda(z)
\en
\end{itemize}
\vspace*{-2ex}
A point $z$ in the effective worldsheet 
represents a 2d block of dimensions  $\Lambda^{-1}\times \Lambda^{-1}$.
The measure $\rho(\lambda(z)) \, d\lambda(z)$  summarizes the
$\lambda$-fluctuations inside the block
that have been integrated out,
hence \emph{a priori} meaning {\it from the earlier}
or {\it from what has gone before}.

The \emph{a priori} measure $\rho(\lambda)d\lambda$ evolves under the 
2d-RG.
It diffuses in $\mathcal{M}$ because of the 
$\lambda$-fluctuations.
At the same time
the 
$\lambda^{i}$ are flowing
along $\beta^{i}(\lambda)$
towards the $\beta=0$ submanifold.
So $\rho(\lambda)d\lambda $ evolves under a driven 
diffusion process.
Taking $d\lambda$ to be the metric volume element,
so $\rho(\lambda)$ is a function on $\mathcal{M}$,
the driven diffusion equation is
\eq
\Lambda \frac{\partial}{\partial \Lambda}\, \rho(\lambda)
= \nabla_{i}\left(g_{\mathrm{str}}^{2} G^{ij} 
\partial_{j} + \beta^{i}\right) \rho(\lambda)
\en
Integrating out the $\lambda$-fluctuations 
up to $\Lambda^{-1}$
drives $\rho(\lambda)d\lambda$
to the equilibrium  measure
$$
 \rho(\lambda) d\lambda\rightarrow 
e^{-g_{\mathrm{str}}^{-2}S(\lambda) }d\lambda 
\qquad
\text{where}
\quad
\beta^{i} = G^{ij}\partial_{j}S
$$
The $\lambda^{i}$ are the space-time field modes with 
UV cutoff $L$
so the \emph{a priori} measure 
$\rho(\lambda)d\lambda$ is the functional integral of 
an effective QFT($L$) with classical action 
$g_{\mathrm{str}}^{-2}S(\lambda)$.

The $\lambda$-model produces
an effective 2d-QFT and
an effective \emph{a priori} measure $\rho(\lambda) d\lambda$
at every 2d distance $\Lambda^{-1}\ll\mu^{-1}$.
Thus the $\lambda$-model produces
an effective S-matrix($L$)
and an effective QFT($L$)
at every distance $L\,{\gg}\, 1$.
The consistency conditions
{\bf C1, C2, C3}
are automatically satisfied.
The S-matrix RG acts on S-matrix($L$) by design.
The QFT RG acts on QFT($L$) because of 
the decoupling of the  effectively irrelevant 
$\lambda^{i}(\Lambda)$
in the 2d-RG.
The agreement between S-matrix($L$) and QFT($L$)
on scattering amplitudes at scales ${\sim} L$
is guaranteed because the scattering amplitudes of S-matrix($L$) 
near the IR cutoff $L$
are given by the 2d correlation functions of vertex operators
near the 2d UV cutoff $\Lambda^{-1}$
which are determined by the \emph{a priori} measure
$\rho(\lambda)d\lambda$
which is QFT($L$).

\subsection{Possible non-canonical degrees of freedom and couplings in SU(2) and SU(3) 
quantum gauge theory}

The $\lambda$-model is a nonperturbative 2d-NLM with possibilities of
nonperturbative semi-classical 2d effects \cite{Friedan2010ua}.
These would be
2d winding modes associated to
$\pi_{1}(\mathcal{M})$ and 2d instantons
associated to $\pi_{2}(\mathcal{M})$
where the target manifold 
$\mathcal{M}$ is the
manifold of space-time fields.
Suppose
the $\lambda$-model produces an effective QFT($L$)
with SU($N$) gauge fields
in four macroscopic space-time dimensions.
Then the mathematical results
\eq
\begin{split}
\pi_{k} \text{ of the manifold of  SU($N$) gauge fields on }
\Reals^{4}\cup\{\infty\}  = \pi_{k+3}(\text{SU}(N))
\\[1ex]
\pi_{4}(\text{SU}(2)) = \Integers_{2}
\qquad
\pi_{5}(\text{SU}(2)) = \Integers_{2}
\qquad
\pi_{4}(\text{SU}(3)) = 0
\qquad
\pi_{5}(\text{SU}(3)) = \Integers
\end{split}
\en
imply that there are 2d winding modes when the gauge group is SU(2) and
2d-instantons when the gauge group is SU(2) or
SU(3).
These nonperturbative semi-classical effects in the $\lambda$-model
offer possibilities of conditional predictions of the form
{\it if the $\lambda$-model produces SM+GR then
it also produces
non-canonical degrees of freedom
from the $\Integers_{2}$ winding mode in
the SU(2) gauge fields and
non-canonical interactions
from the 2d instantons in the 
SU(2) and SU(3) gauge fields.}

\subsection{2d winding modes and 2d instantons}
Let the gauge field $A_{{+}{-}}(u,0)$ be an SU(2) instanton-anti-instanton 
configuration on $\Reals^{4}$
in the limit where one or both of the instanton sizes goes to zero.
The parameter $u$ is the relative orientation.
It lies
in the adjoint representation SU(2)/$\Integers_{2}$.
The remaining moduli of the instanton-anti-instanton pair
are left implicit.
$A_{{+}{-}}(u,0)$ is a solution of the Yang-Mills equation
(which is the 2d-RG equation $\beta=0$).
The nontrivial element of $\pi_{1}=\Integers_{2}$ is the nontrivial closed loop in 
SU(2)/$\Integers_{2}$.
Blowing up the zero-size instanton to a small size $\rho$
gives a four-parameter 
family of gauge fields $A_{{+}{-}}(u,\rho)$ which is an
$\Reals^{4}/\Integers_{2}$ orbifold.  The 2d winding mode is the 
$\Integers_{2}$ twist
field of this orbifold.  The moduli of the $\Integers_{2}$ twist field are the
remaining moduli of the instanton-anti-instanton pair (including
their fermionic zero-modes).

The 2d instantons in the SU(2) and SU(3) gauge fields on $\Reals^{4}$
are also found in zero-size instanton-anti-instanton configurations.
For SU(3) the relative orientation
of a zero-size instanton-anti-instanton pair is parametrized by 
SU(3)/U(1) which has a topologically nontrivial 2-sphere.
For SU(2) the minimal nontrivial 2-sphere is in
the zero-size configurations of two instantons 
and two anti-instantons.

\subsection{Vacuum condensate of SU(2) Yang-Mills flow defects}
\label{sect:vacuumcondensate}

Suppose the target manifold includes SU(2) gauge fields on 
$\Reals^{4}$.
Suppose a portion of the \emph{a priori} measure falls
in the $\Integers_{2}$-odd sector.
Then the $\lambda$-model will contain a gas of $\Integers_{2}$ twist 
fields.
Each $\Integers_{2}$ twist field will pin the worldsheet to an orbifold 
point, a zero-size instanton-anti-instanton pair.
But any $\Integers_{2}$-even cluster of twist fields
will be unstable,
perched high up at twice the instanton 
action without topological protection against being pushed by the 2d-RG
down to the flat SU(2) gauge field.
The 2d-RG acts on the SU(2) gauge fields
as the Yang-Mills flow --- the gradient flow 
of the Yang-Mills action.
For each orbifold fixed point $A_{{+}{-}}(\mathbf{1},0)$
in which the instanton and anti-instanton are aligned,
there is a unique downward trajectory $A_{\mathrm{YMF}}(t)$
that starts from
$A_{{+}{-}}(\mathbf{1},0)$ at $t\,{=}\,{-}\infty$ 
and ends at $t\,{=}\,{+}\infty$ at the flat gauge field.
The instanton and anti-instanton grow together
and annihilate
along the downward trajectory.

A $\Integers_{2}$-even cluster of twist fields will appear in the worldsheet
as a Yang-Mills flow defect operator
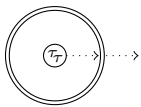
\begin{center}
\begin{tikzpicture}[scale=0.5]
\draw (0,0) circle (1.3);
\draw (0,0) circle (1.2);
\draw[dotted,->] (0.3,+0.0) --  (1.15,+0.0);
\draw[dotted,->] (1.35,+0.0) --  (2.2,+0.0);
\draw (0,0) circle (0.3);
\draw (-0.06,0.05) node{\tiny $\tau$} ;
\draw (0.08,-0.07) node{\tiny $\tau$} ;
\end{tikzpicture}
\end{center}
The core is the $\Integers_{2}$-even cluster of twist fields.
The inner dotted arrow is the slow flow in the $\beta\approx 0$
region near the orbifold fixed point.
The fast part of the flow
is between the solid circles.
The outer dotted arrow is the slow flow near the flat 
gauge field.
The 2d gas of these defects will produce
a vacuum condensate
in the effective QFT($L$).
The downward Yang-Mills trajectory will determine
the properties of the condensate.
The initial and final stages of the trajectory 
can be controlled by perturbing around $\beta=0$.
The intermediate stage can at least be calculated numerically.

\section{To do}

Most urgent
is to determine if the $\lambda$-model does in fact make
conditional predictions of observable non-canonical effects in 
SU(2) and SU(3) gauge theory in 4 dimensions.
This requires
figuring out how to calculate semi-classical corrections to the \emph{a 
priori} measure of the $\lambda$-model
coming from the 2d winding modes and instantons.
Most promising would seem to be
the vacuum condensate
of $\Integers_{2}$ winding modes
for SU(2) gauge fields
sketched in section~\ref{sect:vacuumcondensate} above.
If such conditional predictions can be made and
checked, then there will be compelling motivation for further investigation of the 
$\lambda$-model.
There are many basic questions to investigate.
Most of the technical foundation remains to be built.
But the effort
might not be worthwhile unless and until there is a successful 
conditional prediction.

The $\lambda$-model operates at 2d distances up to $\Lambda^{-1}$.
The space-time distance scale $L$ is
given by $L^{2} = \ln \left( {\Lambda}/{\mu}\right) $.
So the $\lambda$-model builds QFT($L$) from the largest distances
\emph{down} to $L$
(nevertheless ensuring that the QFT RG is satisfied).
This top-down construction of effective QFT
might have useful consequences concerning naturalness
or its lack.

The \emph{a priori} measure $\rho(\lambda)d\lambda$ is a measure on 
the target manifold which is the space of effective 2d-QFTs of the string 
worldsheet.
In the extreme limit $L\rightarrow\infty$, $\Lambda^{-1}\rightarrow 
0$,
only the exactly marginal 2d couplings $\lambda^{i}$ fluctuate. 
The target manifold at $L=\infty$, $\Lambda^{-1}=0$ is the space
of worldsheet 2d conformal field theories
which give idealized asymptotic S-matrices.
The $\lambda$-model dynamically produces a measure 
$\rho(\lambda)d\lambda$ on this
space of idealized classical backgrounds. 
Only if some portion of the measure
concentrates near a background with
large space-time dimensions
will $\rho(\lambda)d\lambda$
take the form of the
functional integral of an effective QFT($L$)
at $L<\infty$.
Exploring the dynamics of such concentration
and the variety of places the 
measure might concentrate
is an intimidating task.
More practical is to
suppose that some part of the measure does concentrate near a large 
4d space-time with space-time fields that include those of the SM+GR,
then try to make predictions
conditional on this assumption.
If these predictions can be checked against experiment, then 
it might be worth asking if the $\lambda$-model
does in fact cause some portion of
$\rho(\lambda)d\lambda$
to concentrate at backgrounds with macroscopic 4d space-times.

Wick rotation is an after-thought.
Space-time euclidean signature   is assumed
so that the 2d-NLM of the worldsheet
will be a well defined effective 2d-QFT.
Then 
only finitely many modes $\lambda^{i}(\Lambda)$ of the space-time fields fluctuate 
and their fluctuations are governed by a positive definite metric 
$G_{ij}(\lambda)$.
So the $\lambda$-model is an {effective} mechanism.
Wick rotation is to be carried out \emph{ad hoc} in the effective 
QFT($L$) and in the effective S-matrix($L$),
if and when the 
\emph{a priori} measure concentrates at a macroscopic space-time.
There is no hint
of an explanation of Wick rotation
in this machinery.

Cosmology might be done by relating
the distance scale $L$ to
the time scale of the cosmological observer,
the early universe being described by
an outgoing scattering state in S-matrix($L$)
and the later universe by a state in QFT($L$).


\addtocontents{toc}{\protect\vspace{0.75ex}
\protect\renewcommand{\protect\cftsecpresnum}{}
\protect\renewcommand{\protect\cftsecaftersnum}{}
\protect\addtolength{\protect\cftsecnumwidth}{-1.5em}
}

\section*{Appendix.  Notes on the line of thought}
\addcontentsline{toc}{section}{\numberline{\hfill \hfill}Appendix.  Notes on the line of thought}
\renewcommand{\thesubsection}{A.\arabic{subsection}}
\setcounter{subsection}{0}


\subsection{Search for a mechanism that produces QFT}
\label{app:devel}

The ideas expressed in this note were developed mainly 
during the period 1977--2002
in the process of searching for and eventually formulating a 
mechanism that would produce quantum field theory.

\subsubsection*{The 2d-RG as a mechanism for space-time physics (1977--79)}

The line of thought began with the renormalization
of the general 2d nonlinear model
\eq
\int  e^{-\int d^{2}z \; g_{\mu\nu}(X) 
\partial X^{\mu} \bar \partial X^{\nu}}
\mathcal{D} X 
\qquad
X(z) \in M
\en
where $g_{\mu\nu}(X)$ 
is a Riemannian metric
on a manifold $M$.
The 2d-RG
\eq
\Lambda \frac{\partial}{\partial \Lambda}\, g_{\mu\nu}(X)  = - 
R_{\mu\nu}(X) +  O(R^{2})
\en
drives the 2d-NLM to a solution of $R_{\mu\nu}=0$.
This was very exciting (at least for me).
The 2d-RG appeared as a \emph{mechanism} that \emph{produces} 
solutions of a GR-like space-time field equation $R_{\mu\nu}=0$.
This suggested the possibility
that a mechanism --- the 2d-RG ---
might actually be the answer to questions like
{\it where does space-time field theory come from?}
or even
{\it where do the laws of physics come from?}
The goal became a {mechanism} that actually {produces} the laws of physics
(instead of the goal of
fundamental principles such as symmetry to constrain the laws of 
physics).

It had become clear by the late 1970s that there are far too many
effective QFTs.  A mechanism was needed that would {produce}
effective QFT more selectively than the QFT renormalization group.
The 2d-RG seemed promising for the purpose since it
at least produced solutions of {classical} field theory.
The goal became a mechanism that produces \emph{quantum} field theory
and that has the 2d-RG as its classical limit.

The general 2d-NLM had two other shortcomings as a mechanism for producing
space-time physics.
First, 
the solutions of $\beta = 0$, the 2d-RG fixed points, have 
unstable directions along which the RG flow diverges from the fixed point 
rather than converges.
Second,
the $\beta=0$ equation $R_{\mu\nu}=0$ is not quite Einstein's equation.

\subsubsection*{The 2d-RG incorporated into string theory (1981--85)}

In the early 1980s
it was realized that the 2d-RG fixed point equation $\beta=0$
(i.e., 2d scale invariance)
is a consistency condition for calculating
the string S-matrix from a worldsheet 2d-QFT.
The string worldsheet is a supersymmetric 2d-QFT containing additional 2d
degrees of freedom besides $X(z)$.
The 2d coupling constants are,
in addition to the space-time metric $g_{\mu\nu}(X)$,
a collection of non-abelian gauge fields,
scalar fields, fermion fields, etc.\ on space-time.
The string worldsheet resolved
the two shortcomings of the basic 2d-NLM.
First, the 2d supersymmetry of the string worldsheet
eliminates the unstable directions 
at the fixed points (the tachyons in the S-matrix).
Second, the worldsheet $\beta=0$ equation
generalizing $R_{\mu\nu}=0$
is a semi-realistic supersymmetric classical field equation that includes GR
and potentially the SM.

In the mid-1980s, several assumptions and mathematical idealizations
became truisms:
\vskip2ex

\begin{enumerate}

\item The string S-matrix was taken to be an asymptotic S-matrix 
without IR cutoff ---
a ``theory of everything''.

\item The string backgrounds were taken to be the conformally 
invariant worldsheet 2d-QFTs
from which such asymptotic string S-matrices are derived --- the exact
worldsheet solutions of $\beta=0$ given by Calabi-Yau manifolds
($R_{\mu\nu}=0$) and generalizations.

\item 
It was \emph{assumed} that the IR physics of string theory 
is the supersymmetric QFT that happens 
to have the same low momentum scattering amplitudes
as the asymptotic string S-matrix.
The string backgrounds  were conflated with those supersymmetric QFTs.

\item It was assumed that there must exist a microscopic QFT or some other kind of 
microscopic
mechanical hamiltonian system from which the string S-matrix is derived.

\end{enumerate}

\subsubsection*{Questions (1987)}

Circa 1987, the key questions seemed to me to be:
\begin{enumerate}
\item 
How does the 2d-RG act in string theory as a \emph{mechanism}?
The fixed point equation
$\beta=0$ is only a consistency condition for the string S-matrix recipe.
\item 
Where does \emph{quantum} field theory come from in string theory?
What produces a functional integral over space-time fields?

\item 
What is the \emph{quantum} string background?
It should be a quantum state in a QFT.

\end{enumerate}

\subsubsection*{The $\lambda$-model (1988-2002)}

The attempt to answer these questions was a long-drawn-out process.
One seed was the idea that the string background is encoded in the 
{local} 2d physics of the worldsheet.
Another seed was the vague notion that nonperturbative effects in string 
theory might come from infinite genus worldsheets.
Eventually, these were combined in the
idea that a froth of small handles would contribute to the
local 2d physics of the worldsheet
and thus to the string background.
This motivated the calculation of the log divergence 
in the contribution of a single small handle
which took the form of a bi-local 
insertion in the worldsheet
(as in section \ref{sect:smallhandlelogdivergence} above).
This log divergence was a strong signal.
The infrared log divergence of the 
scalar field 2-point function plays a 
fundamental role in 2d-QFT.
Thus the idea of setting the 2d coupling constants fluctuating as 2d scalar
fields $\lambda^{i}(z)$ with the 2d-IR log divergence of the  scalar 
field fluctuations 
cancelling the 2d UV log divergences of the small handles.

The essential role of a 2d distance scale $\Lambda^{-1}$
as 2d UV cutoff in the string worldsheet
and as 2d IR cutoff on the $\lambda$-fluctuations
required abandoning the idealized asymptotic string S-matrix ``of everything''
for an effective string S-matrix with IR cutoff $L$
(as in section \ref{sect:effectivestringSmatrix} above).
Integrating out the froth of small handles
became the S-matrix RG.

Recognizing that 
the \emph{a priori} measure $\rho(\lambda)d\lambda $ of the 2d-NLM
would govern the local worldsheet physics
led to 
identifying $\rho(\lambda)d\lambda $
as the functional integral of the effective space-time QFT
that is the quantum string background
(as in sections 
\ref{sect:effective2dcouplingconstants} and 
\ref{sect:productionofeffectiveQFT} above).

At this point the task became to identify semi-classical 2d effects 
in the $\lambda$-model that might lead to checkable predictions.

\subsection{Pragmatism and the S-matrix philosophy}

The S-matrix has been proposed as a formal structure for 
fundamental physics
at several points in history
when QFT has seemed to hit a wall.
The history is recounted in \cite{Cushing1990}.
Heisenberg first proposed using the S-matrix 
as a fundamental formalism in the 1940s
in response to the divergences of perturbative QED
and the difficulty of accounting for
cosmic ray showers.
In the 1960s
the S-matrix bootstrap program was proposed
in response to the plethora of  mesons and baryons
and their strong couplings.
On both occasions QFT overcame its difficulties
and the S-matrix proposals lapsed.
The third occasion was the string S-matrix proposal
of the early 1970s which attracted interest at least in part because of the 
incompatibility between GR and QFT when extrapolated down to the Planck 
length.

Heisenberg's explicit rationale
for using the S-matrix 
was the principle 
that fundamental physics should be expressed in terms of what is actually
observed.
This principle has had notable successes in fundamental theoretical physics.
For example, the route of Bohr and Heisenberg to Matrix Mechanics was 
guided by focussing on observable transitions.
But the principle was not followed literally.
Matrix Mechanics in the end 
described the world by quantum states and transition amplitudes
which are not themselves observable.
Only their absolute squares are observable.
The strategy of focussing on what is observable
led to a formalism, Matrix Mechanics, that reliably
produces observable quantities.
A pragmatic version of the principle might be
\emph{use the minimal formal machinery that is useful to
produce the observable quantities of physics.}
Quantum Mechanics in the form of QFT
is so successful at producing observable quantities
that it can be considered to be
``what is observable'' at distances greater than 
about $(10^{3} 
\text{\small GeV})^{-1}$.

The S-matrix philosophy
proposed replacing Quantum Mechanics and QFT
with an asymptotic S-matrix.
But the asymptotic S-matrix
is an extreme idealization of what is observable.
All the useful work of physics at distances larger than
the elementary  particle frontier
uses Quantum Mechanics or its effective approximation Classical 
Mechanics.
Is it feasible, for example, to describe the behavior of a galaxy
in terms of an S-matrix?
If fundamental physics is formulated as an S-matrix,
what produces effective QFT and effective Quantum Mechanics and effective 
Classical Mechanics?

On the other hand, a pragmatic version of the S-matrix philosophy 
does seem reasonable.
Scattering amplitudes describe what is observable {at short 
distances} where `short' is relative to the size of the observer.
An effective S-matrix with an IR cutoff $L$ is a practical formulation of
what we can actually observe at
distances smaller than the 
limit of our best hamiltonian quantum mechanical model.

From this point of view string theory is interesting not because
it offers an S-matrix ``theory of everything''
but rather because
it is a way to construct 
a self-consistent S-matrix for short distance physics
without requiring a short distance
QFT.
A short distance S-matrix that does not require a short distance QFT
is entirely suitable
in a pragmatic version of the S-matrix philosophy.
The only way to
construct an S-matrix
before string theory
was to derive it from a QFT.
But that does not mean that every
S-matrix and in particular the string S-matrix must be derived from a microscopic QFT.

The pragmatist philospher C.~S.~Peirce (a contemporary of 
Ernst Mach)
proposed that the symbolic tools of science
take their significance
from the work that they do.
(This might well be a selective, idiosyncratic reading of Peirce.)
A pragmatic strategy is to shape
the formalism of fundamental physics
for the work it needs to perform.
The pragmatic view argues against
pursuit of mathematical beauty, 
against pursuit of beautiful fundamental principles,
against attempting to extrapolate to an
absolutely fundamental theory based on absolutely fundamental
principles.
A successful fundamental theory may eventually be based on beautiful 
principles and formulated in beautiful mathematics.
But there is no telling how far away that is
or in what direction.
There is no telling in advance
which mathematically beautiful forms
will prove useful for fundamental physics.
Meanwhile, a practical strategy is to try to make
incremental improvements in the formalism of fundamental 
physics that can actually do useful work in describing the
fundamental physics of the real world.

%
%
\section*{Acknowledgments}
A draft of this note was presented in the  Quantum Gravity Seminar
of the Perimeter Institute.
I thank the members of the Quantum Gravity Group for their comments
and hospitality.
I also thank E.~Rabinovici, 
G.~Moore, and I.~Nidaiev
for comments and questions.
This work was supported by the New High Energy Theory Center
of Rutgers University.

\bibliographystyle{ytphys}
\bibliography{Literature}

\end{document}